\documentclass[conference,10pt]{IEEEtran}
\usepackage{amsmath,amsfonts}
\usepackage{algorithmic}
\usepackage{algorithm}
\usepackage{array}
\usepackage[caption=false,font=normalsize,labelfont=sf,textfont=sf]{subfig}
\usepackage{textcomp}
\usepackage{stfloats}
\usepackage{url}
\usepackage{verbatim}
\usepackage{graphicx}
\usepackage{cite}
\usepackage{multirow}
\usepackage{arydshln}
\usepackage{xcolor}
\usepackage[0]{editing}
\usepackage{acronym}
\usepackage{balance}
\def\BibTeX{{\rm B\kern-.05em{\sc i\kern-.025em b}\kern-.08em
T\kern-.2em\lower.7ex\hbox{E}\kern-.125emX}}
\IEEEoverridecommandlockouts

\begin{document}

\title{Downlink Beamforming Design for NOMA Using Convolutional Neural Networks}

\author{Chentong~Li\textsuperscript{*},
        Saeed~Mohammadzadeh\textsuperscript{*},
    Kanapathippillai Cumanan\textsuperscript{*},
    Octavia A. Dobre\textsuperscript{\dag},\\
    \IEEEauthorblockA{\textsuperscript{*}School of Physics Engineering and Technology - University of York, York, United Kingdom \\
\textsuperscript{\dag}Department of Electrical and Computer Engineering, Memorial University, St. John’s NL A1B, Canada \\
$ \rm  cl2215@york.ac.uk\textsuperscript{*} $,
$ \rm saeed.mohammadzadeh@york.ac.uk\textsuperscript{*}$, 
$ \rm kanapathippillai.cumanan@york.ac.uk\textsuperscript{*}$,
$ \rm odobre@mun.ca\textsuperscript{\dag}$} % <-this % stops a space
\thanks{The work of S. Mohammadzadeh and K. Cumanan were supported by the UK Engineering and Physical Sciences Research Council (EPSRC) under grant number EP/X01309X/1.}% <-this % stops a space
\thanks{The work of O. A. Dobre was supported in part by the Canada Research Chairs Program CRC-2022-00187 and the NSERC Discovery grant RGPIN-2019-04123.}
}
\maketitle
% \author{
% \IEEEauthorblockN{1\textsuperscript{st} Chentong Li}
% \and
% \IEEEauthorblockN{2\textsuperscript{nd} Saeed Mohammadzadeh}
% \and
% \IEEEauthorblockN{3\textsuperscript{rd} Kanapathippillai Cumanan}

% %Chentong~Li, 
% %      Saeed~Mohammadzadeh,~\IEEEmembership{Member,~IEEE,}  
% %   and~Kanapathippillai~Cumanan,~\IEEEmembership{Senior~Member,~IEEE,} \vspace{-0.75em}
%    }
        % <-this % stops a space
%\thanks{This paper was produced by the IEEE Publication Technology Group. They are in Piscataway, NJ.}% <-this % stops a space
%\thanks{Manuscript received April 19, 2021; revised August 16, 2021.}}



\begin{abstract}

Non-orthogonal multiple access (NOMA) and beamforming are well-established techniques for enabling massive connectivity in future wireless networks. However, many optimal beamforming solutions rely on highly complex iterative algorithms and optimization methods, resulting in an increase in computational burden and latency, making them less suitable for delay-sensitive applications and services. To address these challenges, we propose an effective convolutional neural network (CNN)-based approach for beamforming design in downlink NOMA systems to solve the transmit power minimization problem. The proposed method utilizes two representations of channel state information as input features to produce normalized beamforming vectors. Simulation results show that the CNN-based solution closely approximates the optimal label performance while significantly reducing computational time compared to conventional high-complexity algorithms, enhancing its practicality for real-time applications.

%referred to as TCNN and FCNN, where the FCNN input matrix performs better than the TCNN input matrix
%through supervised learning. In this framework, we develop a CNN to address the problem of minimizing total transmit power in a NOMA system. The problem is to ensure that each user meets its minimum Signal-to-Interference-plus-Noise Ratio (SINR) requirement, where using Second-Order Cone Programming (SOCP) is employed to solve the non-convex problem, which is then used to generate training and testing data. We consider two input matrix which named TCNN and FCNN, Simulation results demonstrate that the proposed CNN model can achieve near-optimal solutions.

%Using this framework we construct beamforming neural networks for the minimize the total transmit power problem in a NOMA system. This problem should guarantees that each user meets its minimum signal-to-interference-plus-noise ratio (SINR) requirement, where second-order cone programming (SOCP) is utilized to derive the optimal solution and generate training and testing data. Simulation results demonstrate that CNN can achieve near-optimal solutions.

\end{abstract}


\section{Introduction}

Future wireless communication systems will form a hyper-connected society with unpredictable massive numbers of users and devices. Higher data speeds, reduced latency, increased dependability, extensive connection, and increased energy economy are the motivating factors behind these systems\cite{3_1}. Traditional orthogonal multiple access (OMA) technologies will not be able to meet the rapidly increasing demands of communication users, especially with the increase in the Internet of Things (IoT) devices\cite{3_2}, which pose a substantial challenge in ensuring massive connectivity in 6G and beyond. To address these challenges and meet such unprecedented requirements, non-orthogonal multiple access (NOMA) has emerged as a viable solution by introducing additional degrees of freedom (DoF) in the power domain, enabling non-orthogonal radio resource sharing among users simultaneously \cite{3_5, 3_2}.

Compared to the traditional OMA, which allocates separate time or frequency resources to individual users, NOMA enables users to share these resources simultaneously using different power levels, known as power-domain NOMA\cite{10_7,3_16}. In other words, the base station (BS) in a downlink NOMA system sends a superposition signal to each user with different power levels. Users with better channel conditions or those closer to the BS first decode the signal transmitted to users with worse channel conditions or those farther from the BS, subtract that signal from the received signal, and then decode their signal. Meanwhile, the user with the worst channel condition or farther from the BS treats signals intended for other users as interference and decodes their signal directly. This decoding process for users is known as successive interference cancellation (SIC). To facilitate a successful implementation of SIC and maintain fairness, users with poorer conditions are allocated higher power levels to mitigate interference from other users\cite{3_17, 3_1}.

%Stronger channel condition user will first decode the signal of the poorer channel user, subtract its signal from the received signal, and then decode its own signal. Meanwhile, the user with the poorer channel treats the signal intended to the other user as interference to directly decode its own signal. The decoding process at the users with stronger channel conditions is known as successive interference cancellation (SIC). To facilitate a successful SIC implementations and maintain fairness, users with poorer conditions are allocated higher power levels to mitigate interference from other users\cite{I5, NOMA5G}.

At the same time, integrating beamforming techniques into NOMA systems has been proven to be advantageous, as beamforming enhances signal reception and transmission by directing signals toward the intended users, thus improving the quality of the received signals\cite{lin2023beamforming, inbeam, mohammadzadeh2022robust}. In \cite{22_3}, a beamforming technique is proposed to minimize transmit power and improve network energy efficiency while using NOMA to mitigate interference, thus reducing total transmit power. Furthermore, the work in \cite{wang2022joint} proposes a three-step resource allocation framework developed to optimize beamforming in NOMA systems. However, these optimization processes, which depend on mathematical models or toolboxes, can become complex and lead to significant computational delays as the number of users in the wireless network increases.

%However, traditional beamforming optimization in NOMA systems, such as minimizing the total transmit power while achieving each user's signal-to-interference-and-noise ratio (SINR) problem, utilizes mathematical models and the optimal toolbox to achieve target performance metrics and ensure quality of service. Nevertheless, the complexity of the optimization procedure can cause significant delays in signal transmission as more users join the wireless network. 

%In \cite{mmwavenoma}, a cross-entropy-based beamforming design is introduced for mmWave-NOMA systems to reduce interference and enhance performance. 

Hence, deep learning-based techniques, especially the convolutional neural network (CNN), have been seen as a potential solution to these highly complex problems in mathematical modeling, since it could extract significant features to produce the intended output automatically\cite{13_16}. The power allocation problem with NOMA is addressed by the authors in \cite{13_2}, where a CNN is created with channel data as input, while its outputs are the power. Similarly, in \cite{13_1}, a CNN is proposed to generate uplink power, followed by an algorithm to compute downlink power suitable for real-time applications. Furthermore, in \cite{13_12}, deep CNNs are employed to directly generate beamforming vectors for each user, enhancing performance gains and facilitating large-scale deployment in underwater communication networks. 
%In \cite{powerliang}, a deep learning-based model is used to predict transmit power with higher accuracy.  since they allow the system to learn from channel estimation or other data and extract significant features to produce the intended output automatically

Motivated by the aforementioned work, this paper proposes a CNN-based beamforming method to address the transmit power minimization problem in downlink NOMA systems. The model is trained to map the channel information to the corresponding beamforming vectors, eliminating the need for iterative optimization during inference. Once trained, the model requires only the channel as input to generate beamforming vectors, thereby significantly reducing computational complexity and execution time compared to conventional optimization and iteration based approaches. This efficiency makes the proposed method well suited for real-time and practical scenarios.

%framework and in comparison to iterative techniques, reduces complexity and latency by using channel estimates as the input to generate the normalized beamforming vectors from the trained CNN framework. This approach reduces the complexity associated with conventional beamforming designs that rely on optimization toolboxes. 

\textit{Notations:}
We use lowercase and uppercase boldface letters for vectors and matrices, respectively. $\mathbb{E} \{\cdot\}$ stands for the statistical expectation. The set of real numbers is represented by the symbol $\mathbb{R}$, while the complex numbers are denoted by the symbol $\mathbb{C}$. $(\cdot)^H$ denotes the conjugate transpose of the vectors or matrix. The Euclidean norm of a vector is represented by the symbol $\|\cdot \|_2$. $\Re(\cdot)$ and $\Im(\cdot)$ are the real and imaginary parts of a complex number, respectively. 

\section{System Model and Problem Formulation}

We consider a downlink transmission of the NOMA system, where the BS equipped with $N$ antennas serves $K$ single-antenna users, indexed by the set $\mathcal{K} = \{1, \cdots, K\}$. It is assumed that the BS has perfect channel state information (CSI). Therefore, the signal received by each user $k$ is given as:
\begin{equation}
y_k= \mathbf{h}_k^H \mathbf{x} + n_k, \; \forall k \in \mathcal{K}, 
\end{equation}
where $\mathbf{h}_k \in \mathbb{C}^{N \times 1}$ represents the channel vectors between the BS and the user $k$. The BS simultaneously transmits a superimposed signal of the form $ \mathbf{x} = \sum_{k\in \mathcal{K}} \sqrt{p_k}\mathbf{u}_k s_k$, where $p_k$ denotes the transmitted power allocated to the user $k$. $\mathbf{u}_k \in \mathbb{C}^{N \times 1}$ is the beamforming vector designed for the user $k$, where it has $ \|\mathbf{u}_k\|_2=1$, and $s_k$ is the information signal intended for the user $k$ (assuming that \(\mathbb{E}\{|s_k|^2\} = 1\)). The noise term $n_k$ is modeled as a zero-mean circularly symmetric complex Gaussian random variable with variance $\sigma^2$. 

%And for ease of understanding, we define $\mathbf{w}_k = \sqrt{p_k} \mathbf{u}_k$, where it has $||\mathbf{w}_k||_2 = \sqrt{p_k}$. 

% The downlink (DL) signal-to-interference and noise ratio (SINR) of the user \( k \) can be expressed as:
% \begin{equation}
% \text{SINR}_k^{\text{DL}}=\frac{ p_k |\mathbf{h}_k^H \mathbf{u}_k|^2}{\sum_{i = 1, i\neq k}^{K} p_i |\mathbf{h}_k^H \mathbf{u}_i|^2 +\sigma_k^2}. \label{SINRo}
% \end{equation}

%$\mathbf{w}_k s_k$ to all users, where $s_k$ denotes the information symbol intended for user $k$, satisfying $\mathbb{E} \{|s_k|^2\} = 1$, and $\mathbf{w}_k \in \mathbb{C}^{N \times 1}$ is the corresponding beamforming vector. 

% \begin{figure*}[t]
% \centerline{\includegraphics[width=0.6\textwidth]{NOMA_SIC.png}}
% \caption{A basic downlink NOMA model}
% \label{fig_noma_model}
% \end{figure*}

%following the principle that users are ordered according to their channels,

Users are ordered based on their channel strength, such that $\|\mathbf{h}_1\|_2 \le \|\mathbf{h}_2\|_2 \le \cdots \le \|\mathbf{h}_K\|_2$. For the successful implementation of SIC on users, it is crucial to allocate higher power levels to users with weaker channel conditions. This ensures that their signals are sufficiently strong to be accurately distinguished and decoded. Subsequently, these decoded signals can be subtracted from the received signal in users with better channel conditions, facilitating effective interference cancellation\cite{1_2}. In other words, the user $k$ can decode and subtract the signals of the first $( k-1 )$ users. This can be achieved by ensuring compliance with the following constraints, which must be met to guarantee the desired performance \cite{1_1}:
\begin{equation} \label{sm4} 
\begin{aligned}
    p_1|\mathbf{h}_k^H \mathbf{u}_1|^2 \geq \dots \geq p_k|\mathbf{h}_k^H \mathbf{u}_k|^2 \geq 
     \dots \geq p_K|\mathbf{h}_k^H \mathbf{u}_K|^2, \\\;\forall k\in \mathcal{K}
\end{aligned}
\end{equation}

% |\mathbf{h}_k^H \mathbf{w}_1|^2 \geq \ldots \geq |\mathbf{h}_k^H \mathbf{w}_{k-1}|^2 \geq |\mathbf{h}_k^H \mathbf{w}_k|^2 \geq |\mathbf{h}_k^H \mathbf{w}_{k+1}|^2\geq
%     \\  \ldots \geq |\mathbf{h}_k^H \mathbf{w}_K|^2, \;\forall k\in \mathcal{K}.

In order to measure the quality of service for the user $k$, let us define the signal-to-interference-and-noise ratio (SINR) as follows:
\begin{equation}\label{SINR_ori}
\text{SINR}_k=\frac{p_k|\mathbf{h}_k^H \mathbf{u}_k| ^2}{\sum\limits_{i=k+1} ^ K p_i|\mathbf{h}_k^H \mathbf{u}_i| ^2 +\sigma^2}.
\end{equation}
To facilitate the solution, we introduce a new variable defined as $\mathbf{w}_k = \sqrt{p_k} \mathbf{u}_k$. This reformulation allows the original SINR expression in \eqref{SINR_ori} to be rewritten as follows:
\begin{equation}\label{SINR}
\text{SINR}_k=\frac{|\mathbf{h}_k^H \mathbf{w}_k| ^2}{\sum\limits_{i=k+1} ^ K |\mathbf{h}_k^H \mathbf{w}_i| ^2 +\sigma^2}.
\end{equation}

Hence, the power minimization problem is defined as:
\begin{subequations} \label{pmyuan}
\begin{align}
     & \min _{\mathbf{w}_k\in \mathbb {C}^{N\times 1}}  \sum_{k=1}^{K} \|\mathbf{w}_k \|_2^2     \label{pmyuan1}
   \\ &\text{s.t. }  \frac{|\mathbf{h}_k^H \mathbf{w}_k |^2}{\sum_{i=k+1}^{K}| \mathbf{h}_k^H \mathbf{w}_i| ^2 +\sigma^2} \ge \gamma_k^{\text{min}}  ,\;\forall k \in \mathcal{K},   \label{pmyuan2}
\end{align}
\end{subequations}
where $\gamma_k^{\text{min}}$ represent the minimum SINR threshold.
%and $\|\mathbf{w}_k \|_2^2 $ represents the transmit power assigned to the $k^\text{th}$ user.

% Hence, the received signal at user \( k \), after performing SIC and removing the signals intended for the first \( k-1 \) users, is given by:
% \begin{equation}
%     y_k = \mathbf{h}_k^H \mathbf{w}_k s_k + \sum_{i=k+1}^K \mathbf{h}_k^H \mathbf{w}_i s_i + n_k,\label{rec1}
% \end{equation}
% where $n_k$ implies additive white Gaussian noise with variance $\sigma^2$ and zero-mean circular symmetry. Under these circumstances, the signal-to-interference-and-noise ratio (SINR) for the user \( k \), after applying SIC, is expressed as follows:
% \begin{equation} 
% \text{SINR}_k=\frac{|\mathbf{h}_k^H \mathbf{w}_k |^2}{\sum_{i=k+1}^{K}| \mathbf{h}_k^H \mathbf{w}_i| ^2 +\sigma^2}, \label{sm3} 
% \end{equation}

The original power minimization problem formulated in \eqref{pmyuan} is inherently non-convex due to the non-convex nature of the SINR constraints in \eqref{pmyuan2}. To address this challenge, we reformulate the problem using second-order cone programming, which transforms the non-convex formulation into a convex optimization problem. This reformulation enables the computation of the optimal solution efficiently \cite{1_2, 1_1}.

%By reorganizing the constraint and taking the square root of \eqref{pmyuan2}, the original non-convex constraints can be reformulated into second-order cone (SOC) and linear constraints as follows:

In this design, a phase rotation is applied to the beamforming vectors without affecting the SINR and still achieves the same solutions. This is due to the fact that the SINR values depend on the magnitude of the $\mathbf{h}_k^H \mathbf{w}_k$ terms and not on the phase of $\mathbf{h}_k^H \mathbf{w}_k$\cite{13_17}. In other words, it allows us to assume that $\mathbf{h}_k^H \mathbf{w}_k$ contains only the real part, effectively treating the imaginary part as zero, which, in turn, makes the square root of $| \mathbf{h}_k^H \mathbf{w}_k|^2$ well defined \cite{inbeam,bjornson2013optimal}. Thus, the original non-convex constraints can be transformed into a second-order cone (SOC) and linear constraints by applying the square root to \eqref{pmyuan2}, as shown below:
% \begin{subequations}
\begin{align} \label{eq7}
   \gamma_k^{\text{min}}(\sum_{i=k+1}^{K} |\mathbf{h}_k^H \mathbf{w}_i|^2 + \sigma^2) \le |\mathbf{h}_k^H \mathbf{w}_k|^2  %\label{eq6}
  \nonumber \\ \; \Longleftrightarrow
    \begin{cases}\sqrt{\gamma_k^{\text{min}}}
    \begin{Vmatrix} |\mathbf{h}_k^H \mathbf{w}_{k+1}|
    \\ \vdots
    \\ |\mathbf{h}_k^H \mathbf{w}_K|
    \\ \sigma
    \end{Vmatrix} \le |\mathbf{h}_k^H \mathbf{w}_k|, 
    \\ \Im(\mathbf{h}_k^H \mathbf{w}_k)=0, \quad \forall k \in \mathcal{K}.
    \end{cases} 
\end{align}
% \end{subequations}
%\com{Are you sure about the part I mentioned by green color? I believe it should be $\mathbf{w}_i$. Also, in the below equation. Am I wrong?}

Therefore, the optimization problem in \eqref{pmyuan} can be written in an easier and more tractable format using \eqref{eq7}, allowing us to reformulate it as follows:
\begin{equation}\label{socp}
\begin{aligned} 
    &\min _{\mathbf{w}_k\in \mathbb {C}^{N\times 1}}  \sum_{k =1}^{K} ||\mathbf{w}_k||_2^2  
 \\  & \text{s.t. } \;\;\;\;\;\;
 \begin{cases}\sqrt{\gamma_k^{\text{min}}}
    \begin{Vmatrix} |\mathbf{h}_k^H \mathbf{w}_{k+1}|
    \\ \vdots
    \\|\mathbf{h}_k^H \mathbf{w}_K|
    \\ \sigma
    \end{Vmatrix} \le |\mathbf{h}_k^H \mathbf{w}_k|,
    \\ \Im(\mathbf{h}_k^H \mathbf{w}_k)=0, \quad \forall k \in \mathcal{K}.
    \end{cases}  
\end{aligned}
\end{equation}

Although some optimization toolboxes can efficiently solve the problem in \eqref{socp}, their computational time may be prohibitive when fast and reliable service is essential in the future. Hence, we introduce a CNN-based method, which considers the channel as an input and generates the beamforming vector $\mathbf{u}_k$, to achieve a near-optimal solution and to facilitate practical applicability. Note that to generate labeled input and output pairs for CNN training, we solve the optimization problem in \eqref{socp} with the optimization toolbox, such as the CVX toolbox. 

Furthermore, to verify whether CNN-generated beamforming vectors $\mathbf{u}_k$ satisfy the SINR requirements and minimize total transmit power, we consider the following evaluation approach\cite{13_1}. Given the original expression of SINR in \eqref{SINR_ori}, it is necessary to determine the corresponding transmit power $p_k$ to calculate the SINR for each user. Accordingly, the downlink power allocation $\mathbf{p}$ can be obtained as follows:
\begin{equation}
     \mathbf{p} = \sigma^2 \mathbf{\Psi}^{-1} \mathbf{1},
\end{equation}
where $\mathbf{1} = [1,\dots,1]^T$, and $\mathbf{\Psi} \in \mathbb{C}^{K \times K}$ is given as: 
\begin{equation}
     [\mathbf{\Psi}]_{k i}= 
     \begin{cases}\frac{1}{\gamma_k^{\text{min}}}\left|\mathbf{h}_k^H \mathbf{u}_k\right|^2, & \text { if } k = i, \\ 
    -\left|\mathbf{h}_k^H \mathbf{u}_k\right|^2, & \text { if } k < i , \\
     0, & \text { else. }
     \end{cases}
\end{equation}

\section{The CNN-based beamforming method}

For CNN-based methods, we propose two different input matrix formats and reconstruct the channel to ensure compatibility with CNN model, thus achieving more efficient feature extraction and learning. We then provide a detailed explanation of the architecture and functionality of each layer within the CNN. Finally, we describe the training and testing configurations, along with the implementation details of the proposed network in MATLAB deep learning toolbox.

%This CNN includes an input layer followed by several convolutional blocks, pooling layers, fully connected layers, activation function layer and a regression layer. A typical feature extraction process in CNNs is a convolution layer with nonlinear activation followed by a pooling layer. Convolution layers perform convolution with multiple learnable filters in parallel, and results are fed into nonlinear activation functions to generate feature maps. Then, a pooling function reduces spatial dimension by downsampling the feature maps.  Consequently, the output matrix, which represents the beamforming vectors, is structured to
%Each block consists of a convolutional layer followed by a batch normalization layer and an activation function layer.
%In order to create a feature map, the convolutional layer performs convolutions using various random filters simultaneously. The output is turned into a nonlinear activation function.

\subsection{Proposed CNN Framework}

%\com{You didn't mention 2D before. It is good to mention it at the end of the introduction} \com{As I know, we don't do any channel, so we need to be careful about channel estimation.}

\begin{figure}[!]
\centerline{\includegraphics[width=0.47\textwidth]{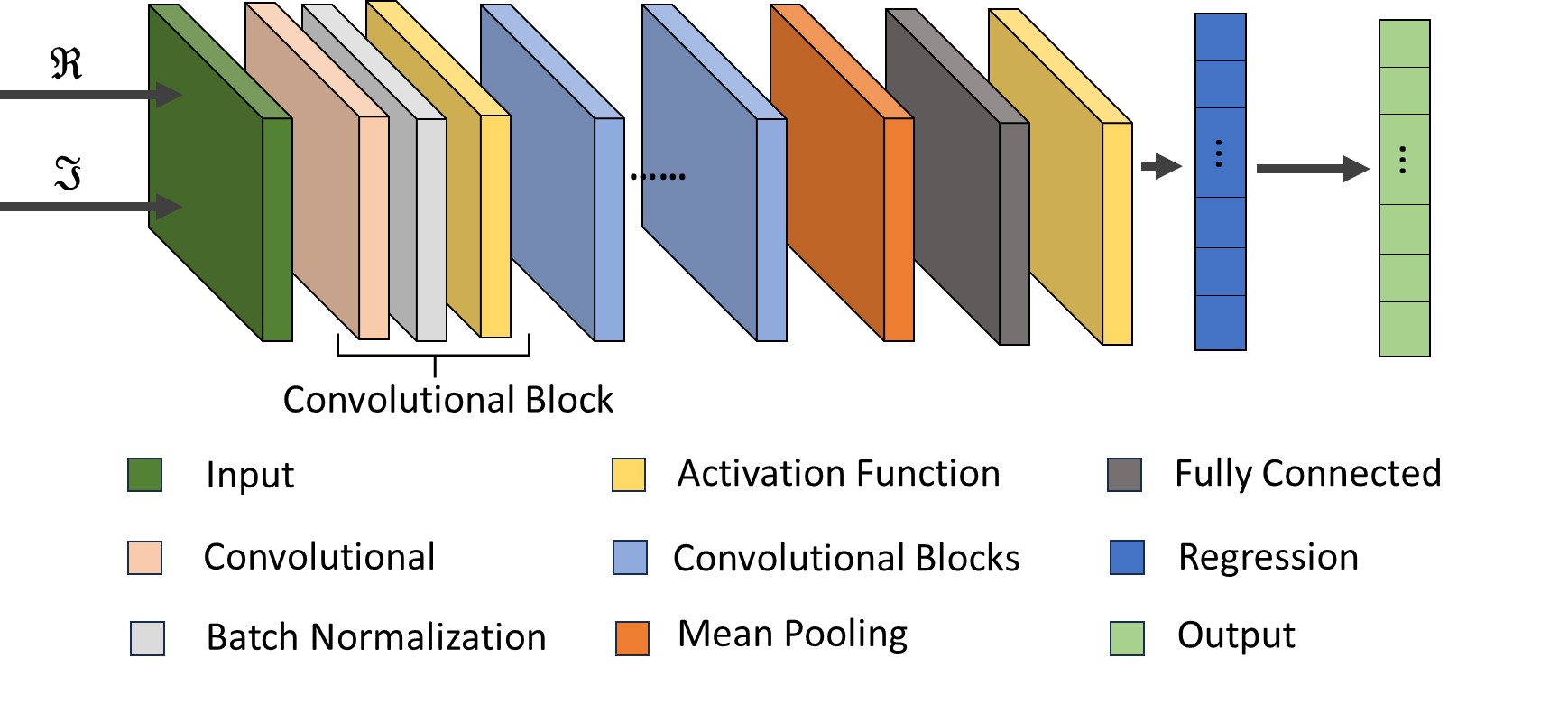}}
\caption{The architecture of the proposed CNN }
\label{Basic_model}
\end{figure}

The architecture of the proposed CNN model is illustrated in Fig.~\ref{Basic_model}. The network begins with an input layer, followed by a sequence of convolutional blocks. Each block consists of a convolutional layer, a batch normalization layer, and an activation function. After multiple such blocks, the network includes a mean pooling layer and a fully connected layer. Finally, a regression layer produces the CNN output. 

Since both channel and beamforming data consist of the complex values (real and imaginary components), and due to limitations in the MATLAB deep learning toolbox, which primarily supports real valued inputs, it is necessary to transform these datasets into alternative real valued formats for network training. Note that we discuss the parameters in the MATLAB toolbox and the size of the input matrix for this layer in Section~\ref{train_test}. 

%The proposed CNN framework takes the channel matrix in the frequency domain as input.  prevent the direct handling of complex numbers as input and output

The first layer of the proposed CNN architecture is the input layer, where the channel is used as label input. Here, we introduce two different input matrix formats, each designed to reorganize the channel in a specific way and then rebuild it into matrix form. For the first input matrix method, the channel $\mathbf{h}= [\mathbf{h}_1^T, \dots, \mathbf{h}_K^T]$ is separated into its real and imaginary components as follows:
\begin{equation} \label{formold}
\begin{aligned}
    \text{TCNN:}\; & \Re(\mathbf{h}) = [\text{real}(\mathbf{h}_1^T), \dots, \text{real}(\mathbf{h}_K^T)] \in \mathbb{R}^{ 1 \times (NK)} ,  \\ & \Im(\mathbf{h}) = [\text{imag}(\mathbf{h}_1^T), \dots,  \text{imag}(\mathbf{h}_K^T)] \in \mathbb{R}^{ 1 \times (NK)},
\end{aligned}
\end{equation}
where we refer to this model as the TCNN, where the complex channel undergoes an \textit{I/Q transformation} as described in \cite{mohammadzadeh2022robust}. Specifically, the input to the convolutional layer is reshaped as $[\Re(\mathbf{h});\Im(\mathbf{h})] \in \mathbb{R}^{2 \times NK}$. However, this transformation treats each user's channel independently, which can limit the ability of the model to capture spatial or inter-user relationships. As the complexity of the output data increases, effectively modeling the dependencies between users becomes increasingly important to improve the CNN prediction accuracy.
%This approach is widely utilized and ensures compatibility with the Matlab toolbox. \com{Have you defined this? or referred to a reference?}
%which has the format below:
% \begin{table}[ht]
% \centering
%     \begin{tabular}{|c|c|c|c|}
%     \hline
%          \text{real}($\mathbf{h}_1^T$) & \text{real}($\mathbf{h}_2^T$) & $\cdots$ & \text{real}($\mathbf{h}_K^T$)\\ 
%     \hline
%          \text{imag}($\mathbf{h}_1^T$) & \text{imag}($\mathbf{h}_2^T$) & $\cdots$ & \text{imag}($\mathbf{h}_K^T$) \\
%     \hline
%     \end{tabular}
%     %\label{FORMOLD}
% \end{table}   
%\flag{This approach is widely utilized and ensures compatibility with the Matlab toolbox. However, in situations where the complexity of output data increases, such as in multi-antenna and multi-user systems, it becomes important to establish interconnections among individual user channels within the input data structure. }\com{Maybe we can write this statement in the simulation section?}

Therefore, we propose to reshape the channel into an alternative format to enhance the interdependence between channel elements and thus improve the reliability of the CNN output. We construct the channel matrix as $\mathbf{H} = [\mathbf{h}_1, \dots, \mathbf{h}_K] \in \mathbb{C}^{N \times K}$, and extract the real and imaginary components as follows: 
\begin{equation}\label{formnew}
\begin{aligned}
   \text{FCNN:}\; & \Re(\mathbf{H}) =  \text{real}(\mathbf{H}) \in \mathbb{R}^{N \times K},
    \\& \Im(\mathbf{H})= \text{imag}(\mathbf{H}) \in \mathbb{R}^{N \times K},
\end{aligned}
\end{equation}
where we call the FCNN. Then we reshape it into a new input matrix, which is given as $ [\Re(\mathbf{H}),-\Im(\mathbf{H}); \Im(\mathbf{H}), \Re(\mathbf{H})] \in \mathbb{R}^{2N \times 2K}$. This approach ensures that more components can interact through convolution, allowing the CNN to extract more information from the input matrix.

Both input formats restructure the channel to suit the training requirements of the CNN better. However, when considering the NOMA scenario, where channels are ordered according to their quality, there is an inherent correlation within the channel coefficients. The input format in \eqref{formnew} is specifically designed to enhance and preserve these inter-channel relationships, providing the CNN with a better opportunity to learn the underlying structure across input elements. Furthermore, \eqref{formnew} employs a square matrix structure, which aligns with the square filter sizes used in the convolution and mean pooling layers. This design not only retains more spatial information but also enables more effective feature extraction compared to the simpler format in \eqref{formold}.

% In here, we define a normalized beamforming vector as $\mathbf{u}_k = \frac{\mathbf{w}_k}{\|\mathbf{w}_k \|}$, note that it satisfies the condition $ \| \mathbf{u}_k\|_2 = 1$ for each user $k$. 

For the output label data, we follow a similar reshaping strategy as used for the input data. The beamforming matrix is denoted by $\mathbf{U} = [\mathbf{u}_1, \mathbf{u}_2, \dots, \mathbf{u}_K] \in \mathbb{C}^{N \times K}$, where each $\mathbf{u}_k \in \mathbb{C}^{N \times 1}$ represents the beamforming vector corresponding to user $k$. For compatibility with the CNN’s regression output, each complex beamforming vector is transformed into a real-valued linear format. Specifically, the real and imaginary parts are separated as $\Re(\mathbf{u}_k) \in \mathbb{R}^{N \times 1}$ and $\Im(\mathbf{u}_k) \in \mathbb{R}^{N \times 1}$, respectively. The final label format is constructed by stacking these components, resulting in a structured real-valued vector suitable for training.
\begin{equation} \label{formoutput}
\begin{aligned}
   & \mathbf{u} = [\Re(\mathbf{u}_1); \Im(\mathbf{u}_1); \Re(\mathbf{u}_2); \Im(\mathbf{u}_2); \\
   & \;\;\;\;\;\;\;\;\;\;\; \dots, \Re(\mathbf{u}_K); \Im(\mathbf{u}_K)] \in \mathbb{R}^{{(2 N K )}\times 1}.  
\end{aligned}
\end{equation}

%Compared to the input format in \eqref{formold}, the format proposed in \eqref{formnew} demonstrates better performance. 

\textit{Architecture of the proposed framework}:
In the proposed framework, the input layer is followed by four complex convolutional blocks. For all input matrix formats, each convolutional layer employs a stride of 1 and zero padding to preserve spatial dimensions. Each layer utilizes 64 kernels of size 3 to extract critical features from the input. To promote stable and consistent feature distributions, batch normalization is applied after each convolutional layer, which accelerates convergence and enhances overall learning efficiency during training.

%The convolutional layer is the key component of convolutional blocks \cite{inmatlab}. However, the input data for each layer in a neural network can change over time during training, which slows down the learning process and makes it more difficult for the network to converge efficiently.
 
Next, the activation function layer in convolutional blocks gives the network nonlinearity, which helps it recognize complex patterns in the input. In contrast to the commonly used rectified linear unit (ReLU) function, which can be expressed as $\text{ReLu} = \max(0,x)$, we employ the leaky ReLU function. This function enables a small, non-zero gradient for negative input values and is given as follows:
\begin{equation}
    \text{Leaky ReLU} (x) = \left \{
    \begin{aligned}
         &x ,   &\; x\ge 0, \\
         &0.01x ,& \; x < 0.
    \end{aligned}
    \right.
\end{equation}

%make the network more robust to translations and distortions of the input data by summarizing local information and 
Following the four convolutional blocks, a mean pooling layer with a kernel size of 3, zero padding, and a stride of 1 is applied to reduce the dimensionality and computational complexity of the subsequent layers. The output is then passed through a fully connected layer, which transforms the feature map into an output matching the size of the labeled data. Finally, a hyperbolic tangent (Tanh) activation function is applied to map the real-valued outputs to the range $[-1, 1]$, ensuring consistency with the normalized label format \cite{inmatlab}. And the form is given as:
\begin{equation}
   \text{Tanh}(x) = \frac{\sinh(x)}{\cosh(x)} = \frac{e^x - e^{-x}}{e^x + e^{-x}}.
\end{equation}

The final regression layer serves as the output of the network, designed to predict continuous values rather than discrete classes. In this work, the root mean square error (RMSE) is adopted as the loss function to guide the training process. Based on the proposed CNN architecture, the model is trained using channel data as input and the corresponding beamforming vectors as output labels, targeting the transmit power minimization problem. These labels are generated using the optimization based methodology previously described, employing a CVX toolbox to solve the underlying optimization problem.

%Additionally, contrast the most efficient solution with the CNN model with two input matrices. Based on this model and facilitate a comparative analysis between these two different input matrix methodologies.

\subsection{Training and Testing} \label{train_test}
To train the proposed CNN model, a total of 20,000 data samples are generated for training, and an additional 5,000 samples are reserved for testing. The network is trained over 100 epochs using a mini batch size of 200. To enhance generalization, the training data is randomly shuffled at the beginning of each epoch, and 20\% of the training set is allocated for validation. 
Each convolutional block in the CNN follows a consistent structure and parameter configuration throughout the network. The training process utilizes the Adam optimizer \cite{kingma2014adam} to update the model parameters, with the RMSE employed as the loss function. The initial learning rate is set at 0.01 and is reduced by a factor of $\rho = 0.5$ after 50 epochs, that is, halfway through the training, resulting in a new learning rate of 0.005. This schedule helps stabilize convergence and prevent overfitting during the later stages of training.

The proposed CNN solution was run on a laptop equipped with an 11th Gen. Intel i5-1145G7 CPU and 16 GB of RAM and implemented in MATLAB R2022a using the deep learning toolbox, with the network architecture reshaped to suit the toolbox's training requirements.

\section{Simulation Results}

%Note that the input data consists of channel estimations in TCNN and FCNN to ensure compatibility with the Matlab toolbox, as shown in \eqref{formold} and \eqref{formnew}, while the labeled output data, the beamforming matrix $\mathbf{U}$, follow the form shown in \eqref{formoutput}. 

We consider a downlink transmission of a four-antenna ($N=4$) NOMA system in BS, which supports $K=3$ single antenna users. The communication channel is characterized by Rayleigh fading. To ensure fairness among users, all users have the same target SINR. The noise variance is given as $\sigma^2=0.1$. We compare the results obtained from CNN-based input methods with those obtained from the labeled data generated using an optimization toolbox. In addition, we evaluate the performance of two conventional beamforming techniques, maximum ratio combining (MRC) and zero-forcing (ZF), and compare their performance with the labeled result and the CNN-based method.

%In the power minimization problem, our aim is to minimize the overall transmit power for each NOMA user while still meeting the SINR threshold. 

\begin{figure}[!]
\centerline{\includegraphics[width=0.5\textwidth]{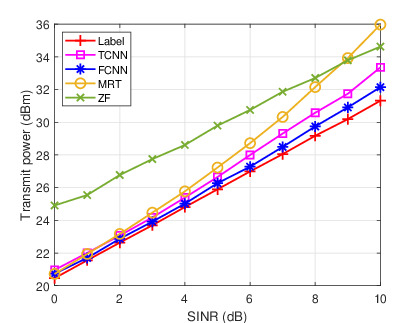}}
    \caption{Transmit power performance versus different SINR thresholds}
\label{fig_socp}
\end{figure}

\begin{figure}[!]
\centerline{\includegraphics[width=0.5\textwidth]{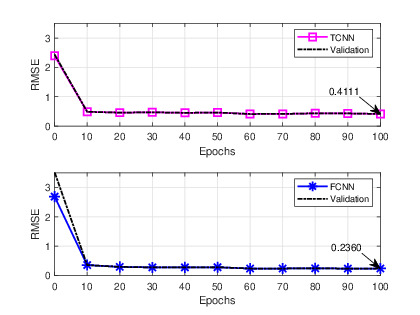}}
\caption{Learning curve of the TCNN and FCNN methods with Adam optimizer}
\label{fig_rmse}
\end{figure}

Fig.~\ref{fig_socp} illustrates the transmit power consumption for varying SINR requirements, comparing with different methods, and using a test dataset of 5,000 samples, the mean total transmit power consumption is evaluated and analyzed. It illustrates that the CNN-based beamforming methods consistently achieve a near-label solution compared to other methods. As SINR increases, CNN-based methods demonstrate more consistent performance across varying SINR levels, whereas the performance gap between MRC, ZF, and the optimal solution becomes more pronounced. Although MRC and ZF are widely used beamforming techniques, they require a higher transmit power to achieve the target SINR compared to the label-based optimization results. In addition, MRC performs relatively well under low SINR requirements, but becomes inefficient as the SINR increases, due to its high power demand. In contrast, ZF consistently exhibits suboptimal performance throughout the SINR range.
Moreover, it is seen that when TCNN and FCNN are compared, the FCNN method outperforms TCNN, with the performance gap widening as the SINR increases. This indicates that FCNN remains more robust and reliable under higher SINR requirements.

%the disparity between the TCNN and the optimal solution becomes more pronounced. In contrast, the FCNN exhibits a more consistent performance across varying SINR levels, with less variation observed. Notably, as SINR increases, the performance gap between TCNN and FCNN widens, indicating that the format in FCNN remains more robust and reliable under higher SINR requirements.  This limitation makes them less suitable for scenarios involving massive numbers of users.two CNN-based input methods, TCNN and FCNN, with MRC and ZF beamforming methods.

Because the TCNN and FCNN perform better in this problem, to compare the convergence speed and performance of the two input matrices, we analyze the maximum number of epochs and their RMSE metrics. Both input matrices are trained under a SINR threshold of 5$\text{dB}$, using the same training and testing data, as illustrated in Fig.~\ref{fig_rmse}. The results demonstrate that both CNN-based beamforming methods converge toward their validation values in the maximum epochs, indicating that neither of the models exhibits signs of overfitting. Although both methods demonstrate relatively low RMSE within the first 10 epochs, additional training epochs are necessary to allow the network to further adjust its weights and effectively learn the features from the input data. Furthermore, the RMSE comparison at convergence shows that the RMSE of FCNN is lower than that of TCNN, indicating that the performance of FCNN is superior.

% \begin{figure}[!]
% \centerline{\includegraphics[width=0.45\textwidth]{bar_char_0401.eps}}
%     \caption{Computation time of Label, TCNN and FCNN methods}
% \label{fig_time}
% \end{figure}

\begin{figure}[!]
	\centering
\includegraphics[width=0.5\textwidth]{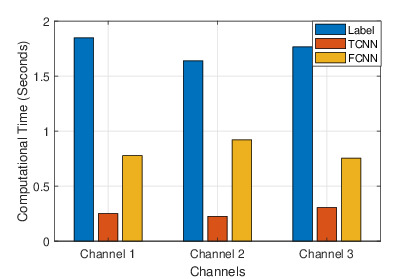}
	\caption{Computation time of label, TCNN, and FCNN methods}
  \label{fig_time}
\end{figure} 

%\begin{table}[!] width=3.2in,height=2.1in
%\centering
 % \caption{Computational Time}
%    \begin{tabular}{|c|c|c|c|c|}
%    \hline
 %    \multirow{2}*{Channels} & SINR threshold & \multirow{2}*{Method} & Time & Total power   \\
%      & (\text{dB}) & & (Seconds) & (\text{dBm})\\
 %   \hline
 %    Channel 1 & 8  & CVX & 1.847046 &  29.0751 \\
%    \hline
%      &  & TCNN & 0.249871 &  30.5703\\
%    \hline
 %     &  & FCNN & 0.776412 &  29.3305 \\
%    \hline
 %    Channel 2 & 9  & CVX & 1.637320 & 29.3233\\ 
%    \hline
%      &  & TCNN & 0.226225 &  31.2536\\ 
 %   \hline
 %     &  & FCNN & 0.919445 & 30.2084\\
%    \hline
 %    Channel 3 & 10  & CVX & 1.764014 & 29.8882\\ 
 %   \hline
 %     &  & TCNN & 0.304301 & 31.5038\\ 
 %   \hline
 %     &  & FCNN & 0.755559 &  30.4429\\
 %   \hline
 %   \end{tabular}
 %   \label{tabletime}
%\end{table}   

To evaluate the time consumption of the proposed methods, we compare the results of the label method with fully trained TCNN and FCNN on three different channels, as summarized in Fig.~\ref{fig_time}. The results demonstrate that fully trained CNN-based beamforming methods are more suitable for real-time applications, as they achieve significantly shorter processing times compared to the label method. This efficiency is attributed to the fact that CNN-based methods use a large number of input data during training to approximate near-optimal solutions. Once trained, the model can generate beamforming results without iterative computations, thus significantly reducing the execution time. Furthermore, although the FCNN requires a longer training duration than the TCNN due to its more complex input matrix, both CNN-based methods substantially outperform the label-based solution in terms of computational efficiency.

%The proposed CNN model is well-suited for real-time applications, offering reduced computational complexity and faster processing times. To evaluate its performance, we compared the results of the CVX-based solution with those of the trained CNN and FCNN models under the same channel estimation and SINR threshold conditions, as summarized in Table~\ref{tabletime}. The result indicates that, for a given channel and SINR threshold, the CVX method requires significantly more processing time, approximately one second longer than the CNN-based approach, which can achieve a fast and near-optimal solution.
%Furthermore, while TCNN uses less time than FCNN, it requires higher total transmit power.

\section{Conclusion}

In this paper, we proposed a CNN-based model to solve a power minimization problem in a downlink NOMA beamforming system. The model is designed to generate beamforming directly using the channel as input. To enhance the performance of the CNN model, we explore two different input matrices: TCNN and FCNN. The result indicates that CNN-based methods outperform other beamforming methods, achieving solutions closer to the label solution method. Furthermore, under a given SINR threshold, the FCNN consistently achieves a near-label solution and lower RMSE compared to the TCNN. The results considering the processing time demonstrate that the CNN-based model significantly reduces computational time while maintaining near-optimal performance.

\balance

\bibliographystyle{IEEEtran}
\bibliography{DLSOCPNOMA}

%\vfill

\end{document}